\documentclass[aps,twocolumn,amsfonts,amsmath,amssymb,prb]{revtex4}
\pdfoutput=1
\usepackage{graphicx}
\usepackage{graphics}
\newcommand*{\revision}[1]{{#1}}

\newcommand*{\revrem}[1]{}

\begin{document}

\title{Computer simulations of ionic liquids at electrochemical interfaces}

\author{C. Merlet$^{1,2}$}
\author{B. Rotenberg$^{1,2}$} 
\author{P.A. Madden$^3$}
\author{M. Salanne$^{1,2,*}$}

\affiliation{$^1$ UPMC Univ Paris 06, CNRS, ESPCI, UMR 7612, PECSA, F-75005 Paris, France}
\affiliation{$^2$ R\'eseau sur le Stockage Electrochimique de l'Energie (RS2E), FR CNRS 3459, France}
\affiliation{$^3$ Department of Materials, University of Oxford, Parks Road, Oxford OX1 3PH, UK}
\affiliation{$^*$ email: mathieu.salanne@upmc.fr}

\begin{abstract}
Ionic liquids are widely used as electrolytes in electrochemical devices. In this context, many experimental and theoretical approaches have been recently developed for characterizing their interface with electrodes. In this perspective article, we review the most recent advances in the field of computer simulations (mainly molecular dynamics). A methodology for simulating electrodes at constant electrical potential is presented. Several types of electrode geometries have been investigated by many groups, in order to model planar, corrugated and porous materials and we summarize the results obtained in terms of the structure of the liquids. This structure governs the quantity of charge which can be stored at the surface of the electrode for a given applied potential, which is the relevant quantity for the highly topical use of ionic liquids in supercapacitors (also known as electrochemical double-layer capacitors). A key feature, which was also shown by atomic force microscopy and surface force apparatus experiments, is the formation of a layered structure for all ionic liquids at the surface of planar electrodes. This organization cannot take place inside nanoporous electrodes, which results in a much better performance for the latter in supercapacitors. The agreement between simulations and electrochemical experiments remains qualitative only though, and we outline future directions which should enhance the predictive power of computer simulations. In the longer term, atomistic simulations will also be applied to the case of electron transfer reactions at the interface, enabling the application to a broader area of problems in electrochemistry, and the few recent works in this field are also commented upon.
\end{abstract}
\maketitle
Ionic liquids are room-temperature molten salts, composed mostly of organic ions for which almost unlimited structural variations are possible~\cite{hallett2011a}.  The scientific and technological importance of ionic liquids derives from a wide-range of applications~\cite{armand2009a,lovelock2012a}. For example they can be used as  solvents \revision{in} the textile industry, as lubricants, or as \revision{heat-transfer} fluids for thermal engines\revision{. D}ue to their \revision{excellent properties as} solvents they may even \revision{enable} emerging technologies in such fields. Here we will focus on the electrochemical applications of ionic liquids only. These can be separated in two main families: electro-deposition on the one hand and energy storage and conversion on the other hand. For the former, \revision{the large electrochemical windows of many} ionic liquids allow for the development of processes that are impossible in water; for example the electroplating of aluminium in order to protect steel from corrosion~\cite{liu2006a}. As for the very-active field of electrochemical storage of energy, many synthesis routes \revision{for novel materials} involve the use of ionic liquids~\cite{barpanda2011a}. They are also used in replacement of  conventional organic solvents as electrolytes in battery~\cite{borgel2009a}, fuel cell or supercapacitor~\cite{liu2010a} devices, allowing for \revision{the exploitation of their chemical stability across} a large electric potential window. 

\begin{figure}[ht]
\begin{center}
 \includegraphics[width=\columnwidth]{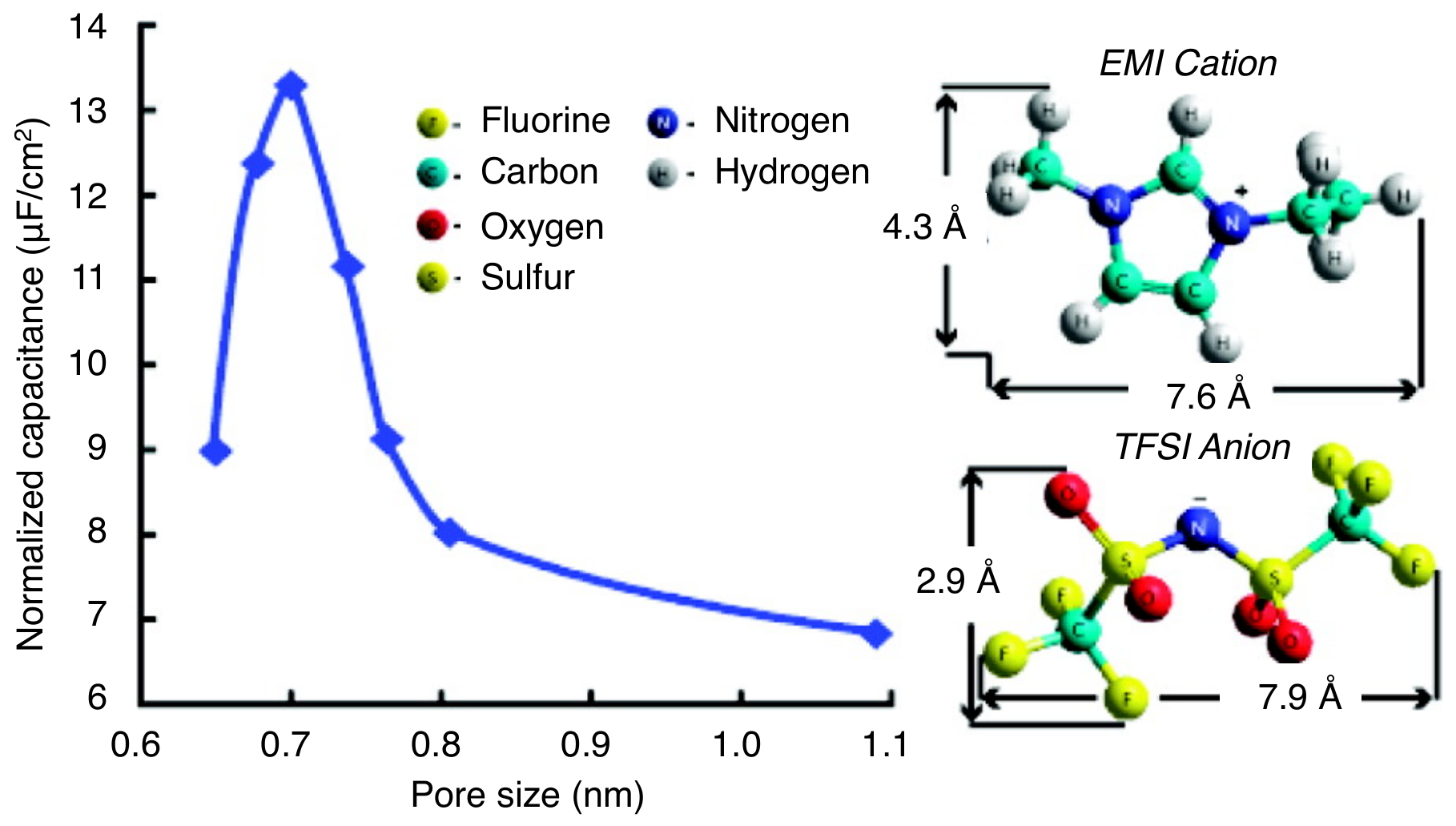}
\end{center}
\caption{``Anomalous'' increase of the capacitance in nanoporous carbons with a ethyl-methylimidazolium-bis(trifluoromethane-sulfonyl)imide ionic liquid electrolyte. Reproduced with permission from reference \cite{largeot2008a}. Copyright 2008 American Chemical Society.}
\label{fig:largeot}
\end{figure}

In particular, it is worth \revision{emphasizing} the case of electrical double layer capacitors (EDLC), which ha\revision{s} attracted much attention in recent years~\cite{simon2008a}. \revision{In these devices, the operating voltage is a crucial parameter, which impacts both the power and energy densities}~\cite{conway-book}\revision{. Many groups have therefore tried to develop ionic liquids with large electrochemical windows~\cite{balducci2007a,krause2011a}. The other target properties are a low viscosity, in order to decrease the charge/discharge time, down to very low temperatures~\cite{lin2011a}, and the capacitance at the ionic liquid/electrode interface.} The discovery of nanoporous electrode materials with enhanced capacitive performances when using an ionic liquid with ion sizes matching the pore size, as illustrated in Figure \ref{fig:largeot}, opens the way for a widespread use of supercapacitors in many contexts where high power electrical output is required~\cite{miller2008a}. As an extension, the behavior of the ionic liquids at charged interfaces even allows \revision{one} to forecast the development of electroactuators~\cite{liu2010b,liu2013a} which could be used as artificial muscles, sensors, and even energy generators in turbulent flows or sea-tide.

The \revision{surge of} experimental results, some of them somewhat unexpected, \revision{which has followed this technical interest and been made possible by new experimental methods,} has spurred the interest of the theoretic\revision{al} community. Despite the under\revision{appreciated} early simulation work of Heyes and Clarke~\cite{heyes1981a}, the interface between electrodes and ionic liquids had barely \revision{been} explored until the last decade. \revision{Since t}hen several studies have focused on the structure of ionic liquids on charged flat surfaces~\cite{lanning2004a,pinilla2007a}, \revision{and shown} that the ion\revision{ densities} exhibit a pronounced oscillatory structure close to the interface. In parallel, the development of a mean-field theory based on the Poisson-Boltzmann lattice-gas model \revision{has shown} that it is mandatory to account for the finite volume occupied by the ions, resulting in a dramatic departure from the Gouy-Chapman law for the \revision{dependence} of the capacitance on potential~\cite{kornyshev2007a}. It is worth noting that the latter work actually preceded experiments, and that numerous studies have reported differential capacitance results in qualitative agreement with the mean-field theory since then~\cite{islam2008a,kruempelmann2010a}.  

\revision{More refined} simulation studies have focused on a better understanding of the existence of particular capacitance-potential curve shapes, such as e.g. the ``bell-shaped'' or ``camel-shaped'' ones. In particular, the roles of the asymmetry of size between the ions or of the charge distribution inside a given ion have been adressed~\cite{fedorov2008a,fedorov2010a,georgi2010a,vatamanu2010a}. It was also shown that for the particular case of the interface between the molten salt LiCl and an aluminium electrode, a potential-induced ordering transition of the adsorbed layer may be at the origin  of a \revision{sharp} peak of the capacitance-potential curve~\cite{pounds2009b,tazi2010a}. Such a situation has also been reported for imidazolium ionic liquids systems in experiments based on {\it in situ} scanning tunneling microscope~\cite{su2009a,su2010a}. Lately, the adsorption on atomically corrugated electrodes was shown to induce a systematic increase in the differential capacitance compared to the case of flat electrodes~\cite{vatamanu2011b,vatamanu2012a,xing2012a}.

The technologically important case of porous electrode has also recently been investigated. Simulations involving slit-like pores~\cite{vatamanu2010a} or carbon nanotubes~\cite{yang2009a,shim2010a} provided a first insight \revision{into}  the structure of ionic liquids in \revision{a} confined environment. \revision{More recently, the importance of the polarization of the (conducting) electrodes by ions in close proximity to them has been appreciated, and shown to play the key role in allowing the ``superionic" state, which accounts for the high degree of charging at the interface inferred from experiments on porous electrodes, to develop.}~\cite{kondrat2011a,kondrat2011b}. The mechanism at the origin of the enhanced capacitance in nanoporous electrodes was then fully understood from molecular dynamics simulations involving realistic \revision{models of} carbon materials and \revision{which took} into account the polarization in an appropriate way~\cite{merlet2012a}. In order to underline the current vigor of the simulation community in this field, we may mention the following example: the existence of a capacitance varying with an oscillatory behavior depending on the pore size was simultaneously reported by three different groups in the same week~\cite{wu2011a,feng2011c,jiang2011a}!

\section{Methods}

Computer simulations involving ionic liquids at the molecular scale employ either the molecular dynamics (MD) or Monte-Carlo (MC) method~\cite{lyndenbell2007a}. In both cases, it is necessary to determine the intramolecular and intermolecular interactions at each step of the simulation. Among the various existing methodologies, the size of the systems involved in electrochemical applications limits the options to the use of classical approaches, where all these interactions are defined through an effective force field. The intramolecular and van der Waals (repulsion + dispersion) terms generally differ from one study to another in the analytic expressions involved (e.g. Lennard-Jones vs. Born-Huggins-Meyer for the van der Waals term) or in the parameterization scheme~\cite{canongialopes2004a,borodin2009a,dommert2012a,canongialopes2012a}, with very little impact on the qualitative results for the structure (the situation is somewhat more complicated for the dynamics, but the latter has barely been addressed in interfacial systems). In some cases, the computational cost is decreased by employing coarse-grained models~\cite{wang2009a,roy2010a}. 

More \revision{significant} are the differences in the treatment of the electrostatic problem \revision{and, in particular, the influence of the presence of the charged and conducting electrode on the effective interactions}. The electrodes may be treated as ideal conductors  and in macroscopic theory the charge $\rho_{\rm ind}({\bf r})$ \revision{induced on the electrode by the applied potential and the charges of the ions in the electrolyte}  is obtained from the Poisson equation under the condition that, for ${\bf r}$ inside the metal,
\begin{equation}
\Psi({\bf r})=\int {\rm d}{\bf r}' \frac{\rho({\bf}{{\bf r}'})}{\mid {\bf r}'-{\bf r}\mid}=\Psi_0
\label{eq:potential}
\end{equation} 
\noindent where $\rho({\bf r})=\rho_{\rm ind}({\bf r})+\rho_{\rm IL}({\bf r})$ and $\rho_{\rm IL}({\bf r})$ is the charge density due to the ionic liquid (represented by a set of point charges located on the atomic sites). $\Psi_0$ is the applied potential, which should be uniform inside a conducting electrode. \revision{To solve} this equation, \revision{which must be done ``on-the-fly" as the ions move,} it is necessary to include specific developments in the simulation codes, and for this reason in most of the simulation studies \revision{a much simpler representation of the ``electrode" was used}. In that case the electrode is either represented by a planar surface with a fixed uniform surface charge density  (i.e. with no explicit atomic sites) or with a set of fixed partial charges located on fixed atomic positions. In the remaining text we will designate such simulations as {\it constant charge} ones and those in which the induced charge is determined self-consistently by solving equation \ref{eq:potential} as {\it constant potential} ones.

\begin{figure*}[ht!]
\begin{center}
 \includegraphics[width=.8\textwidth]{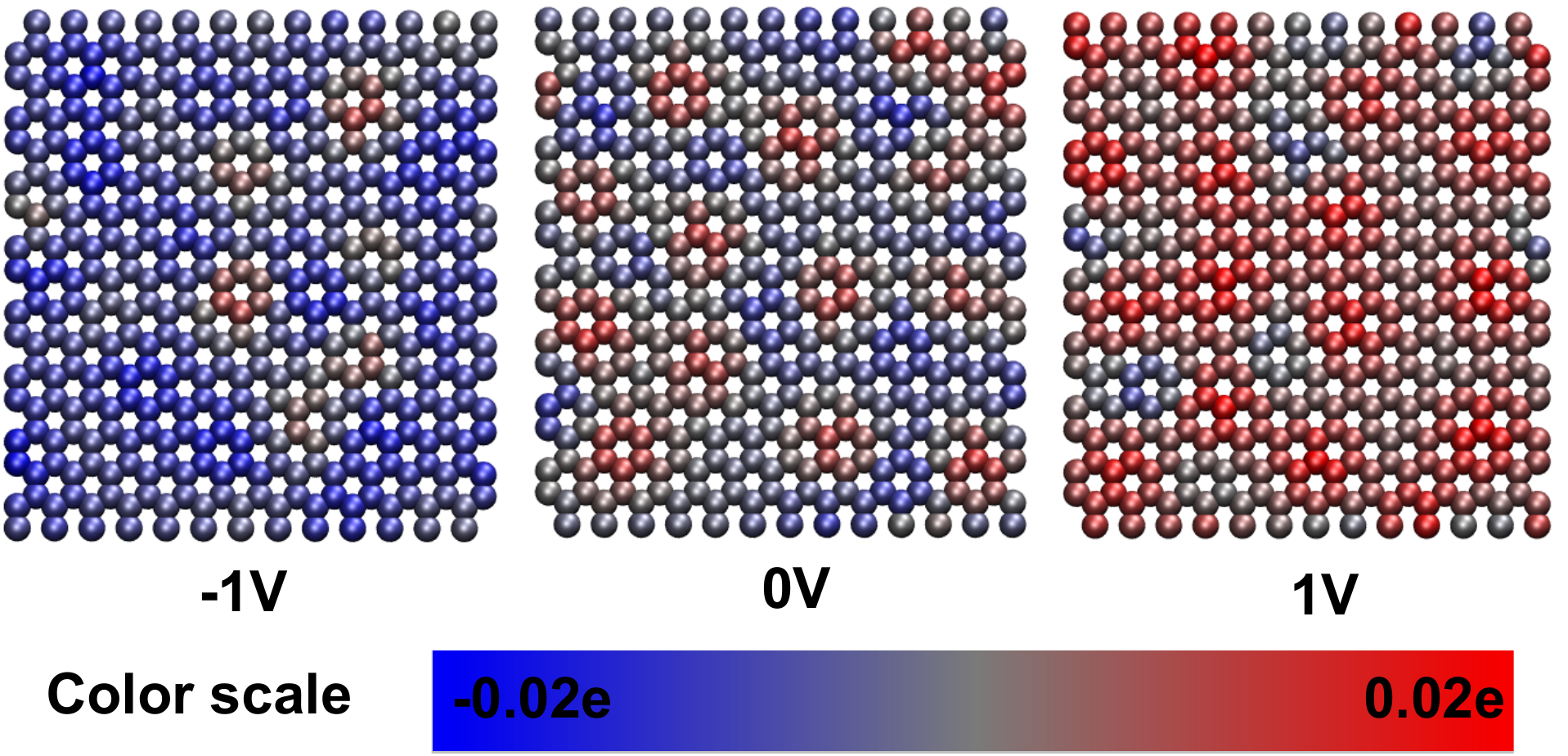}
\end{center}
\caption{Charge on each carbon atom of a planar graphite surface held at different potentials, for instantaneous configurations. }
\label{fig:electrode}
\end{figure*}

In parallel, several methods have been developed in order to perform {\it constant potential} simulations. In particular, Siepmann and Sprik  proposed a model for studying the adsorption of water molecules at a metal surface and on the tip of a model scanning tunneling microscope probe~\cite{siepmann1995a}, which was adapted by Reed {\it et al.} to the case of electrochemical cells~\cite{reed2007a}. In this model, the electrode consists of atoms which can be arranged with a crystalline order or in a disordered way; each atom $j$ carries a Gaussian charge distribution $\rho_j({\bf r})$ which has an integrated charge of $q_j$ and is of fixed width $\eta$:
\begin{equation}
\rho_j({\bf r})=q_j A \exp \left(-\mid {\bf r}-{\bf r}_j\mid^2 \eta^2 \right),
\end{equation}
\noindent where $A=\eta^3 \pi^{3/2}$ is a normalization constant. In our application, the full system consists of two parallel electrodes, which are infinite in the $x$ and $y$ direction, enclosing a set of melt ions. The Coulomb energy, given by,
\begin{equation}
U_c=\frac{1}{2}\int \int {\rm d}{\bf r} {\rm d}{\bf r}' \frac{\rho({\bf}{{\bf r}})\rho({\bf}{{\bf r}'})}{\mid {\bf r}'-{\bf r}\mid},
\label{eq:coulombenergy}
\end{equation}
\noindent has to be expressed through a two-dimensional Ewald summation, for which the correct expressions are provided in reference \cite{gingrich2010a} (note that the use of three-dimensional Ewald summation may lead to artifacts, although no systematic comparison has been undertaken up to now). By combining equations \ref{eq:potential} and \ref{eq:coulombenergy}, we immediately see that the potential experienced by any charge is obtained from the partial derivative of this expression with respect to that charge
\begin{equation}
\Psi_j=\left[ \frac{\partial U_c}{\partial q_j}\right]_{{\{q_{i}\}}_{i\ne j}}.
\end{equation}
\noindent The value of the charge on each electrode atom at each time step in a MD procedure is obtained by requiring that the potential experienced by each charge $j$ in that electrode be equal to the preset electrode potential value, $\Psi_0$, thus satisfying the constant potential condition. This condition is achieved by adding a constraint term, 
\begin{equation}
U_{\rm constraint} = - \sum_{j=1}^M \Psi_0 q_j
\end{equation}
where $M$ is the number of electrode atoms, and by minimizing the total potential energy with respect to all the variable charges simultaneously (note that the two electrodes will have different preset potential values $\Psi_0^{\rm left}$ and $\Psi_0^{\rm right}$, which will be noted $\Psi^+$ and $\Psi^-$ in the following).

A second approach has also been used in the ionic liquid / electrode context, in which the electrode surface is modeled as an equi-potential smooth surface with a net charge~\cite{wu2012a}. Here, the entire electrode surface is effectively modeled as a metal foil separating the ions and a set of fixed charges located inside the electrode. The constant electrical potential is enforced on numerical grid points lying over the electrode surface by solving an auxiliary Laplace equation, following a method proposed in another context by Raghunathan and Aluru~\cite{raghunathan2007a}. This method differs from the Reed {\it et al.} approach by treating the surface as smooth and by replacing the self-consistent calculation of the charges by solving Poisson's equation on a grid.
 

In their grand canonical MC simulation study of a model ionic liquid in a slit-like metallic nanopore, Kondrat and Kornyshev have employed a third method, which consists in modifying the electrostatic interactions between pairs of ions by introducing an effective screening by the metallic wall~\cite{kondrat2011b}. The electrostatic interaction between two ions of respective positions $z_i$ and $z_j$, charges $q_i$ and $q_j$, separated by a lateral distance $R$, is given by
\begin{equation}
U_c^{ij}=\frac{4q_i q_j}{\epsilon L}\sum_{n=1}^\infty \sin (\pi n z_i/L) \sin (\pi n z_j/L) K_0(\pi nR/L)
\end{equation}
\noindent where $\epsilon$ is the dielectric constant of a medium between the plates, $L$ is the pore size and $K_0(x)$ is the zero order modified Bessel function of the second kind. The applied electrostatic potential $\Psi_0$ is introduced by adding a shift in the chemical potential of the ions. This approach is limited to pore geometries for which the screening function can be evaluated analytically, and \revision{only} the first order polarization effects  are taken into account, so that the results are mostly qualitative. 

More recently, another computationally efficient alternative method was proposed by Petersen {\it et al.}, in which the polarization of the electrode by the electrolyte is accounted for by introducing  explicit image charges together with a fluctuating uniform electrode charge located on the electrode surface. The constant potential condition is then enforced by adding a constant uniform charge~\cite{petersen2012a}. The main advantage is that it avoids the need of a self-consistent calculation at each time step, but it presents the drawback of being {\it a priori} limited to the case of electrodes with planar geometries. 

It is also worth noting the work by P\'adua {\it et al.}, who have studied the solvation and stabilization of metallic nanoparticles in ionic liquids, an application in which it is not necessary to control the potential~\cite{pensado2011a,mendonca2012a,mendonca2013a}. In this work, a polarizable model is used for the nanoparticles, which is done in practice by assigning a freely rotating dipole to each atom.

In practice, {\it constant potential} simulations are more easily comparable to experimental data than {\it constant charge} ones, because the conditions are more realistic. \revision{Furthermore, the first-mentioned method above can be applied to electrode surfaces with arbitrary geometries because the charges are determined self-consistently and can vary from site-to-site depending on the local geometry, whereas in {\it constant charge} methods the charges must be preassigned with due regard to the symmetry.} In the remain\revision{der} of the text we will therefore favor the results obtained by using such methods. Nevertheless, constant charge simulations are easier to carry out (no particular modification has to be made in regular simulation packages) and their computational cost is lower (by an order of magnitude). A much larger number of situations have therefore been studied in this framework, and we will also discuss such results when no data from constant potential simulations are available. Note that we have recently shown that\revision{, for simple electrode geometries where the average charge distribution can be predicted,} in general constant charge simulations provide a fair description of the structure of the liquid at the surface of an electrode, and thus of the capacitance. For the dynamics, the situation is much more dramatic:  In particular, upon
\revision{switching on an} applied potential difference, the increase in the temperature, due to
the Joule effect, associated with the creation of an electric current across the cell follows Ohm's
law, while unphysically high temperatures \revision{rapidly develop when the potential change is induced by changing the} charges assigned to each carbon atom.~\cite{merlet2013b} As a consequence, charging kinetics are largely affected by the simulation method, being unrealistically high for constant charge simulations.~\cite{vatamanu2011a,merlet2013b}

\section{Planar electrodes}

\subsection{Structure}
\begin{figure}[ht!]
\begin{center}
 \includegraphics[width=.8\columnwidth]{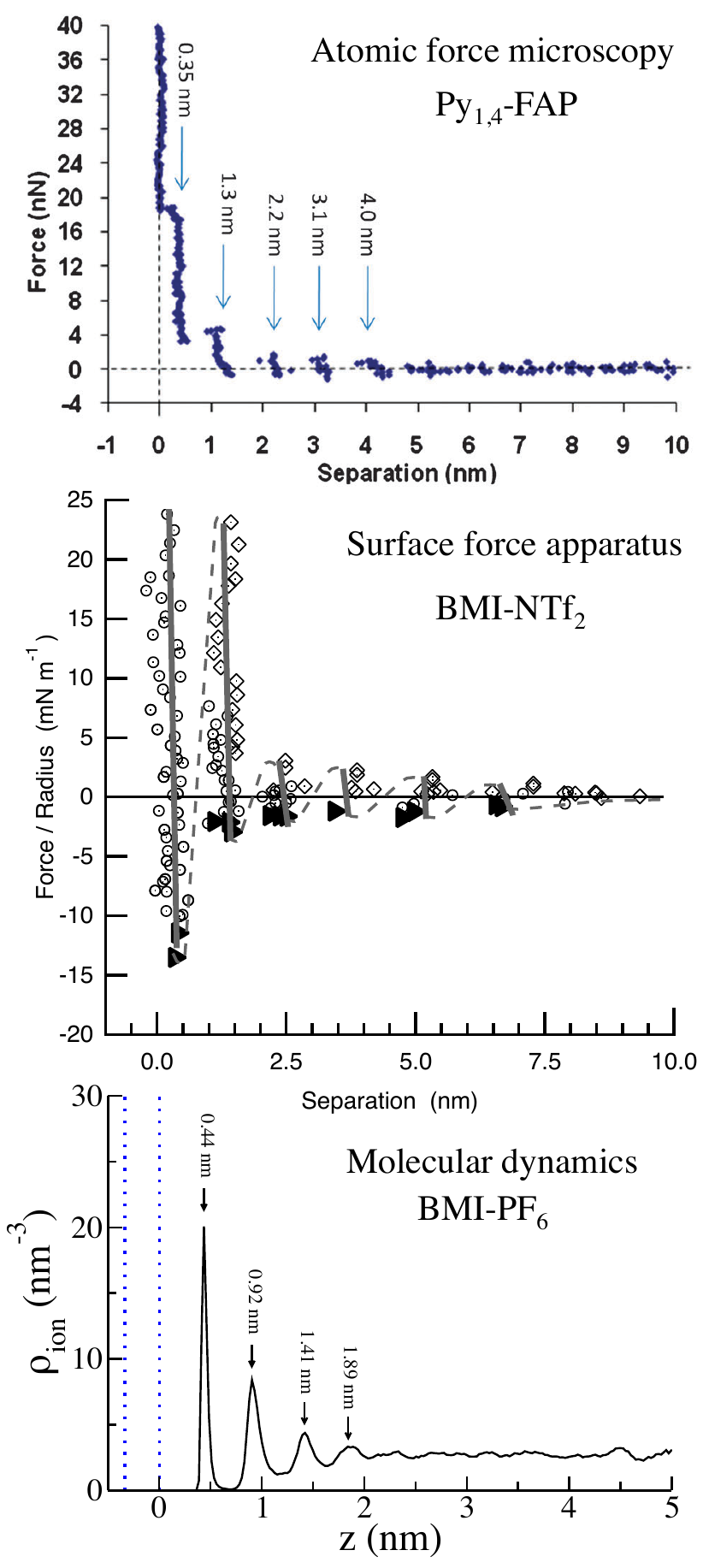}
\end{center}
\caption{Illustration of the layered structure of ionic liquids on planar electrodes. Top: Typical force vs. distance profile for an AFM tip approaching a Au (111) surface in 1-butyl-1-methylpyrrolidinium tris(pentafluoroethyl)trifluorophosphate, at the open circuit potential. Reproduced with permission from reference \cite{atkin2011a}. Middle: Typical force vs. distance profile for two approaching mica surfaces in 1-butyl-3-methylimidazolium bis(trifluoromethylsulfonyl)imide. Reproduced with permission from reference \cite{perkin2011a}. Bottom: Density profiles for the anions extracted from a simulation of an electrochemical cell using 1-butyl-3-methylimidazolium hexafluorophosphate and graphite electrodes (held at a potential difference of 0~V)~\cite{merlet2011a}. }
\label{fig:atkin}
\end{figure}

\revision{The} vast majority of the simulation work  up to now has  been directed toward the understanding of the adsorption of ionic liquids on planar electrodes. In addition to the changes in the simulation method (choice of the force field, of the simulation conditions, of constant charge vs. constant potential, {\it etc}), they vary by the nature of the ionic liquids (molten salts~\cite{heyes1981a,lanning2004a,reed2007a,pounds2009b,tazi2010a,vatamanu2010a}, imidazolium-based ionic liquids~\cite{kislenko2009a,feng2009a,merlet2011a,merlet2012b}, primitive models~\cite{fedorov2008a,fedorov2008b,fedorov2010a,georgi2010a})  and of the electrode (graphite/graphene~\cite{kislenko2009a,merlet2011a,merlet2012b}, smooth surface~\cite{fedorov2008a,fedorov2008b,fedorov2010a,georgi2010a}, generic metal~\cite{vatamanu2010a},LiFePO$_4$~\cite{smith2009a} or solid aluminium~\cite{pounds2009b,tazi2010a}) involved. Nevertheless the main picture of the structure has remained essentially the same since the pioneering work from Heyes and Clarke~\cite{heyes1981a}: The ionic liquid adopts a layered structure, and the structure of the bulk is recovered after several layers. This particular pattern has since then been confirmed by several experimental techniques such as high-energy X-ray reflectivity~\cite{mezger2008a}, AFM~\cite{atkin2007a,hayes2010a,atkin2011a} or surface force apparatus (SFA)~\cite{perkin2010a,perkin2011a,perkin2012a,smith2013a}. Figure \ref{fig:atkin} illustrates this layering as it is observed in AFM~\cite{atkin2011a} (top) and SFA~\cite{perkin2011a} (middle) experiments or in MD simulations~\cite{merlet2011a} (bottom). Similar patterns are observed for all the ionic liquids: In the \revision{illustrated} example, the ionic liquids are composed of \revision{ions with different sizes and shapes, and the nature of the solid surface also changes, which explains the variations in the distances between successive planes in the different cases. Sum Frequency Generation (SFG) is another experimental technique which is able to probe the vibrational properties of molecules at interfaces. In the case of ionic liquids, Baldelli did not deduce from his experiments the presence of several layers of ions at the surface of an electrode~\cite{baldelli2005a,baldelli2008a,baldelli2013a}. It is now possible to determine SFG spectra from molecular dynamics simulations with a high degree of accuracy~\cite{morita2008a,sulpizi2013a}, and the application of such methods to ionic liquids interfaces will certainly help to understand this discrepancy in the future.}

\begin{figure}[ht]
\begin{center}
 \includegraphics[width=\columnwidth]{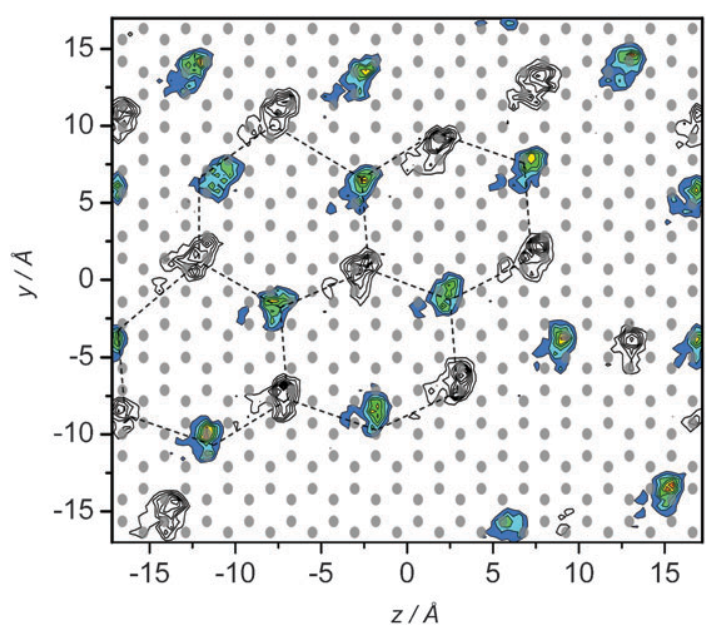}
\end{center}
\caption{Density contour map for ions adsorbed on a graphite surface. Grey
circles indicate surface carbon atoms. The color map corresponds to PF$_6^-$
 anions; the black-and-white contour map corresponds to
imidazolium rings. Reproduced with permission from reference \cite{kislenko2009a}.}
\label{fig:kislenko}
\end{figure}

\revision{Within the plane of} the adsorbed layers, the ionic liquid can also take a noteworthy
organization. Of course, in a given layer, and when the electrode is not
charged, the domina\revision{nt} interactions are \revision{the short-range repulsion,
which controls the packing of ions, and the Coulombic interactions} which occur
between the various ions. A cation is then surrounded by anions and {\it
vice-versa}. Nevertheless {\it in situ} scanning tunnel microscopy (STM) studies
have provided evidence of potential-induced transitions in the adsorbed layers
of ionic liquids, with the formation of ordered structures on the surface of the
electrode~\cite{pan2006a,pan2007a,su2009a,su2010a}. In computer simulations,
Kislenko {\it et al.} have first shown that in pure BMI-PF$_6$, ions are
arranged in a highly defective 2D hexagonal lattice on a neutral graphite
electrode, as illustrated from the density contour maps on figure
\ref{fig:kislenko}.   The ordering of the ions in the first adsorbed layers was also observed in the LiCl // aluminium interface~\cite{pounds2009a}, and it was shown that the structure of the ionic liquid film was commensurate with \revision{that} of the aluminium surface, with different patterning depending on the metallic face oriented towards the liquid~\cite{tazi2010a}.  The question of the orientational ordering of the ionic liquid has also been addressed in numerous computer simulations~\cite{pinilla2007a,dou2011a,mendonca2012a,payal2012a,shimizu2012a,si2012a,lyndenbell2012a}.

Upon charging of the electrode, the ionic liquid keeps a layered structure at the interface. The accumulation of charge on the surface of the metal leads to the formation of additional layers: A segregation of the ionic species is progressively observed~\cite{merlet2011a}, which is at the origin of an {\it overscreening} effect~\cite{lanning2004a,fedorov2008a,fedorov2008b,feng2011a,bazant2011a}. The first adsorbed layer carries a charge which is \revision{larger in magnitude (but opposite in sign) to that on the} electrode,\revision{which is then} counterbalanced in the subsequent layer. This phenomenon extends in the following layers, until the bulk structure is recovered (i.e. up to several nm). This overscreening effect is maximum for low voltage electrodes, and it disappears when high charging r\'egime \revision{is} reached~\cite{fedorov2008a,fedorov2008b} (typically for metal surface charge greater than 16.0~$\mu$C cm$^{-2}$, i.e. \revision{substantially} out of the electrochemical window of any electrolyte, even an ionic liquid~\cite{merlet2011a}!).

\subsection{Capacitance}

\begin{figure}[ht]
\begin{center}
 \includegraphics[width=\columnwidth]{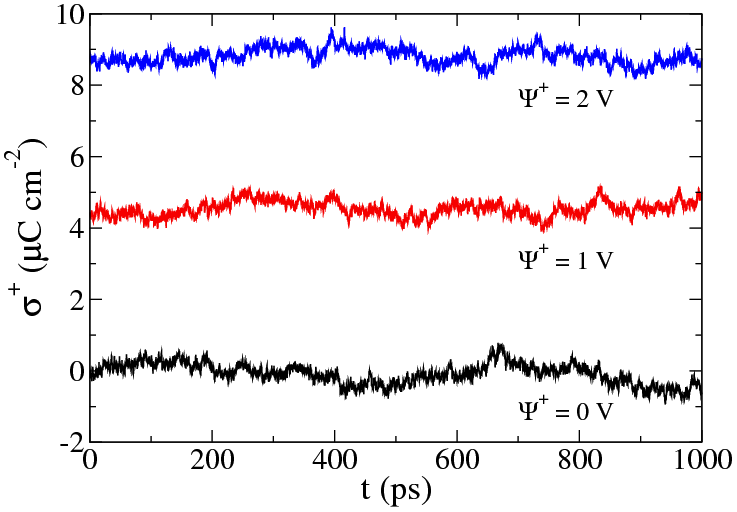}
\end{center}
\caption{Fluctuations of the surface charge for electrodes held at various
positive potentials during molecular dynamics simulations of a system consisting in graphite electrodes and BMI-PF$_6$ ionic liquid~\cite{merlet2011a}. The potential on the other electrode is set to the opposite
value.}
\label{fig:chargefluctuations}
\end{figure}

The \revision{observable property targetted in these} simulations is  the capacitance of the electrochemical cell. One can define either a differential capacitance, which is determined for each electrode by
\begin{equation}
C_{\rm diff}=\frac{\partial \sigma^\pm}{\partial \Delta \Psi},
\end{equation}
\noindent where $\sigma^\pm$ is the average surface charge of the electrode\footnote{The symbol $\pm$ is used for quantities which can be calculated for both the positive and negative electrodes} and $\Delta \Psi=\Psi^\pm-\Psi^{\rm bulk}$ is the potential drop at the electrode/electrolyte interface, 
 or an integral capacitance, given for the full electrochemical cell by
\begin{equation}
C_{\rm int}=\frac{\sigma^{\pm}}{\Psi^+-\Psi^-}.
\end{equation}
\noindent \revision{In} constant charge simulations, $\sigma^+$ and $\sigma^-$ are imposed and the various potential differences involving $\Psi^+$, $\Psi^-$ and $\Psi^{\rm bulk}$ are obtained from the charge density distribution along the normal to the interface:
\begin{equation}  
\Psi(z)=\Psi(z_0)-\frac{1}{\epsilon_0}\int_{z_0}^z {\rm d}z' \int_{-\infty}^{z'}{\rm d}z'' \rho(z'')
\label{eq:poisson}
\end{equation}
\noindent where $\Psi(z_0)$ is an arbitrary integration constant (note that this relation is only valid for planar electrodes). In constant potential simulations, $\Psi^+$ and $\Psi^-$ are imposed and equation \ref{eq:poisson} is also used to determine $\Psi^{\rm bulk}$, but now we have the boundary condition $\Psi(z_0)=\Psi^+$ if $z_0$ is inside the left electrode ($\Psi^-$ in the right electrode). 
In addition, the surface charges are calculated at each time step as explained in the Methods subsection. Typical fluctuations of the positive electrode surface charge along simulation time are shown on figure \ref{fig:chargefluctuations}.

\begin{figure}[ht]
\begin{center}
 \includegraphics[width=\columnwidth]{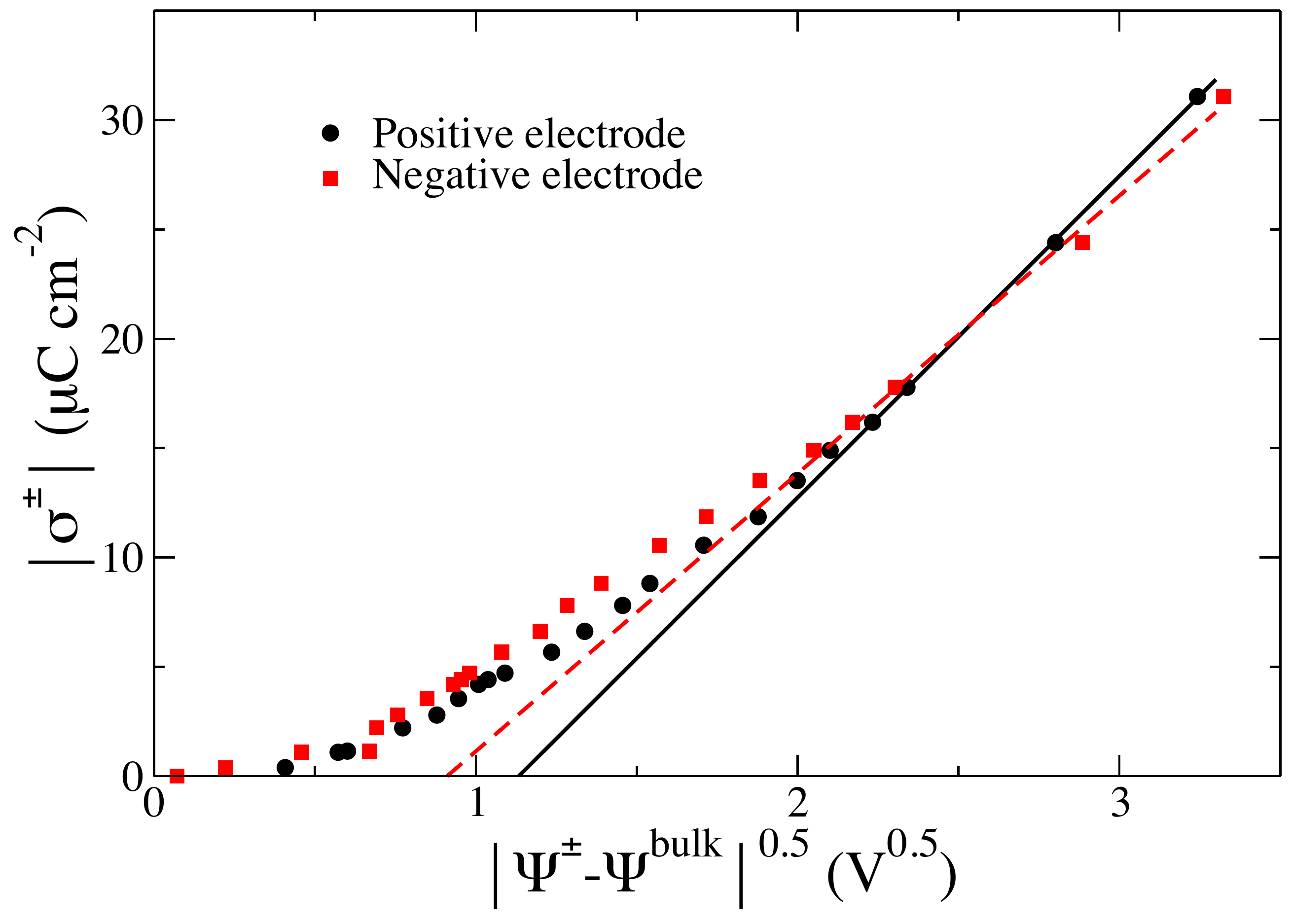}
\end{center}
\caption{Variation of the average surface charge of the positive and negative electrodes w.r.t. the square-root of the potential drop at the interface, for graphite electrodes and BMI-PF$_6$ ionic liquid~\cite{merlet2011a}. The linear correlations indicate the formation of a  lattice saturation r\'egime at high potentials~\cite{kornyshev2007a}.}
\label{fig:sigmavspot}
\end{figure}

A typical $\sigma^\pm =f(\Delta \Psi)$ plot, obtained for an electrochemical cell
made of graphite electrodes and BMI-PF$_6$ ionic liquid, is shown on  figure \ref{fig:sigmavspot}. In our previous work~\cite{merlet2011a},
we proposed linear fits of the data for both the $\Delta \Psi \in [-2 {\rm V}; 0
{\rm V}]$ and $[0 {\rm V}; 2.5 {\rm V}]$ regions and observed that this fit was
not  valid outside the electrochemical stability window. Nevertheless it
was not possible to check the existence of a region where the so-called {\it
lattice saturation} effect occurs, as predicted in the mean field theory
developed by Kornyshev~\cite{kornyshev2007a}. By \revision{extending the
potential range} we can
now assess this point; as shown on figure \ref{fig:sigmavspot},
there is a proportional increase of $\mid \sigma^\pm\mid$ with respect to $\sqrt{\mid \Delta \Psi \mid}$ for potential drops greater than $\approx$~4~V, which is in agreement with the theoretical prediction. In this r\'egime, there is no more overscreening, and on the contrary the ions cannot pack densely enough inside one layer to counterbalance the electrode charge. Comparison with other simulation work seems to indicate that the potential drop for which {\it lattice saturation} effects occur are largely dependent on the nature of the ionic liquid~\cite{georgi2010a,vatamanu2011a,feng2012a}, but the differences in the models and methods between the various studies c\revision{ould} also be  the origin of this variation.

\begin{figure}[ht]
\begin{center}
 \includegraphics[width=\columnwidth]{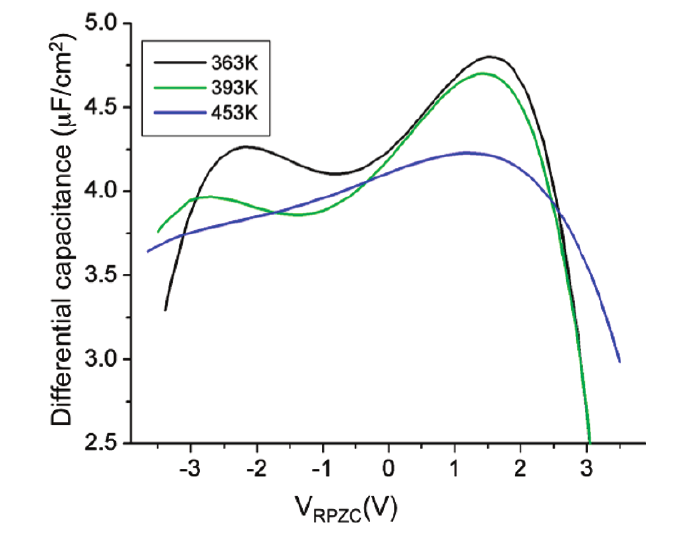}\\
\end{center}
\caption{Differential capacitance as a function of electrode potential
relative to the potential of zero charge (RPZC) for a pyr$_{13}$-TFSI ionic liquid. Reproduced with permission from reference \cite{vatamanu2010b}. Copyright 2010 American Chemical Society.}
\label{fig:capas}
\end{figure}

If we now come back \revision{to potentials within} the electrochemical stability window of the ionic liquids electrolytes, whether there is a general shape for the differential capacitance vs. potential drop (C-V) curves is still an open question. Indeed, many experimental results have been reported following the theoretical work by Kornyshev \revision{which predicted} the existence of ``bell-shaped'' curves. Non trivial shapes (e.g. ``camel-shape'') were measured~\cite{alam2007a,islam2008a,islam2009a,alam2009a,silva2008a}. Nevertheless Lockett {\it et al.} showed that many hysteresis effects occured in these experiments~\cite{lockett2008a}, an observation which was later confirmed by Dr\"uschler {\it et al.}~\cite{druschler2010a}. An important parameter is the structure of the electrode surface, which \revision{is often far from the perfect flat surface assumed in the theory}  since most of the experiments involve either polycr\revision{y}stalline metal surfaces or glassy carbon electrodes~\cite{lockett2010b}. It was also shown that the use of single-frequency measurements could lead to misleading results~\cite{druschler2011a,roling2012a,druschler2012a}.  

Although some of the experimental problems do not arise in computer simulations,
where\revision{,} for example\revision{,} it is much easier to study a perfect monocr\revision{y}stalline surface
than a polycrystalline one, th\revision{ey} also have their drawbacks. The most important
one is that classical modelling is not suitable for the determination of the
potential of zero charge (PZC), which is defined as the value of $\Delta \Psi$ for
which $\sigma^+=\sigma^-=0$ ; of course a value can be extracted from C-V curves
but those cannot be compared to experiments \revision{because it corresponds to the position of the Fermi level of the electrode, an electronic property which can only be caught at the {\it ab initio} level}. {\it Ab initio} molecular dynamics
may provide a solution to this problem in \revision{future} years~\cite{cheng2012a}, but
this technique remains \revision{substantially} too expensive in computational resources \revision{because of the large cells and long relaxation times necessary for the IL interfacial studies} -- for
the moment it is limited to the study of relatively small bulk ionic liquids
samples~\cite{zahn2010a,wendler2012a,salanne2012a}. \revision{The 
surface charges and the relative potentials} fluctuate on long timescales so that their averages are associated with substantial error bars;
for example on figure \ref{fig:chargefluctuations}.  In turn, the method used to extract the C-V
curve is also subject to numerical uncertainties. For example, the linear fits
of the positive and negative branches of the surface charge vs. potential drop
plots can be replaced by polynomial fits as suggested by other
groups~\cite{fedorov2008a,fedorov2008b,vatamanu2011a}.
\revision{We are currently exploring a new route to the differential
capacitance, namely the equilibrium fluctuations of the electrode charge
in constant-potential simulations. The variance of this quantity can indeed be
shown to be related to the differential capacitance~\cite{vanroij2013a}, which could then be
computed without resorting to simulations at different potentials.} 

Typical C-V plots obtained by Vatamanu {\it et al.}\cite{vatamanu2010b} from molecular dynamics simulations of the ionic liquid N-methyl-N-propylpyrrolidinium bis(trifluoromethane)sulfonyl imide (pyr$_{13}$-TFSI) near
a graphite electrode are shown on figure \ref{fig:capas}. The authors showed that the various maxima/minima were due to changes in both the ion
density and in their orientation towards the surface. The shape of the ions and their asymetry therefore play an important role; for example at 363~K and 393~K  the lower value of the maximum of $C_{\rm diff}$ at negative potentials compared to positive potential seems to reflect the relatively high density of the uncharged pyr$_{13}$ propyl tails and ring carbon atoms near the graphite surface for the former, while the latter corresponds to an innermost adsorbed layer of highly charged oxygen atoms of the TFSI anion. 

\section{Curved/rough electrodes}

The study of planar electrodes is important for basic understanding of the
electrical double layer formed with ionic liquid electrolytes. But both at the
\revision{laboratory} and industrial scale, EDLC devices employ active matter
with a somewhat more complicated structure: for example one can find
nanotubes~\cite{alzubaidi2012a}, nano-onions, cluster-assembled nanostructured
carbons~\cite{bettini2013a}, carbide-derived nanoporous
carbons~\cite{gogotsi2003a}\revision{,  etc}. In all these setups, the carbon surface is mostly curved, and it is important to understand how this impacts the structure of the adsorbed ionic liquid and the capacitance, {\it inter alia}.

\begin{figure}[ht]
\begin{center}
 \includegraphics[width=\columnwidth]{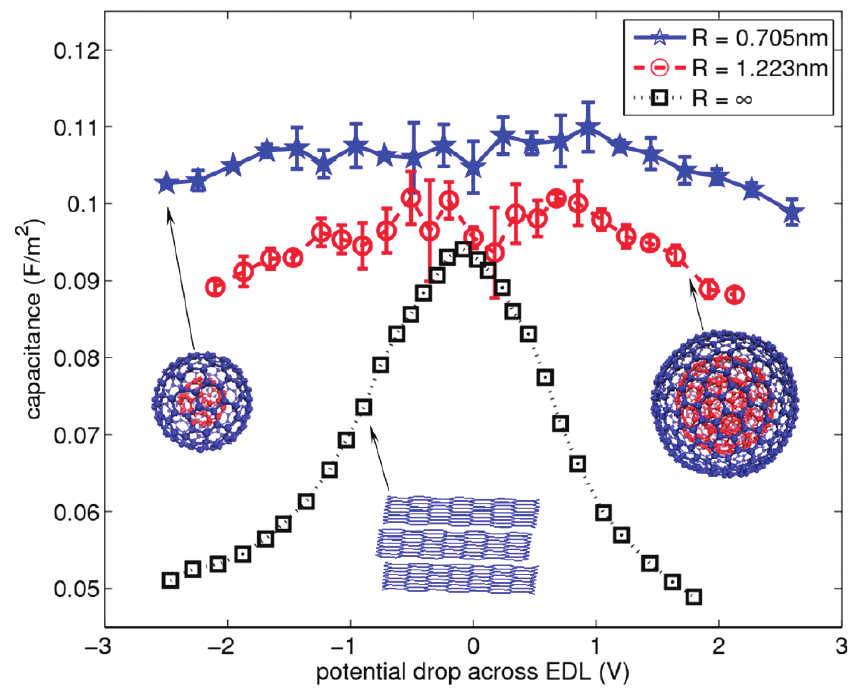}
\end{center}
\caption{Influence of electrode curvature on differential capacitance for an EMI-Tf$_2$N ionic liquid.
$R$ is the radius of an onion-like carbon and is infinity for the planar electrodes. Reproduced with permission from reference \cite{feng2012a}. Copyright 2012 American Chemical Society.}
\label{fig:fromfeng2012a}
\end{figure}

Several groups have therefore tackled this problem using molecular dynamics simulations. On the one hand, Feng {\it et al.} have simulated\revision{,} using a constant charge setup\revision{,} both the outer surface of carbon nanotubes~\cite{feng2011b,feng2013a} and nano-onions~\cite{feng2012a}. In both studies, an overall increase of the capacitance per surface area was observed \revision{compared to a planar surface}. The main result from reference ~\cite{feng2012a} is shown on figure ~\ref{fig:fromfeng2012a}. It reports the variation of the differential capacitance, with respect to the potential drop across each interface (i.e. between the positive electrode and the electrolyte or between the negative electrode and the electrolyte), for a 1-ethyl-3-methylimidazolium bis-(trifluoromethylsylfonyl)imide (EMI-Tf$_2$N) ionic liquid and three different electrode structure\revision{s}: planar graphite and two nano-onions of different radius (0.705~nm and 1.223~nm). The former shows a peak at 0~V, which does not appear in the two other cases. The lattice saturation occurs at relatively small potentials for the planar electrode (-1.0~V on the cathode side and 0.8~V on the anode side) while it does not occur in the investigated potential range for the nanonions. We therefore expect this effect to play an important role in the observed difference. The authors also showed that when the nano-onion becomes smaller (thus increasing the surface curvature), the number of ions accumulated per unit of area of the electrode rises, which leads to an increase of the capacitance (by $\approx$~10~\% across the whole potential range). 

On the other hand, Vatamanu {\it et al.} have investigated rough surfaces using
constant potential simulations~\cite{vatamanu2011b}. To this end, they have
modelled the carbon electrode by using several graphite layers which are stacked
perpendicularly to the interface. They used a\revision{n} ABAB stacking sequence, \revision{which leads to a surface with a regular indentation of 1.43~\AA\ due to the relative positions of the atoms at the edges of the A and B layers}. 
The ionic liquid consists \revision{of} 1-ethyl-3-methylimidazolium cations and
bisfluorosulfonylimide anions. In agreement with the work of Feng {\it et al.},
they showed that a large increase of the capacitance per unit area is obtained, \revision{relative to a planar graphene-like surface}. The shape of the differential capacitance \revision{vs. potential curve} is different, though, since
a peak is obtained for the roughened interface but not for the planar one. Another
important consequence of their study is that even for a simple structure like
graphite, the shape of the capacitance vs. potential plot can vary dramatically
depending on the surface which is facing the liquid, a point which should be
kept in \revision{mind} when comparing simulations when experimental data. This
was also observed in our study of the interface between a LiCl molten salt and
two different faces of aluminum electrodes, \revision{namely} (100) and (110)~\cite{tazi2010a}.

\section{Porous electrodes}

\begin{figure*}[ht!]
\begin{center}
 \includegraphics[width=\textwidth]{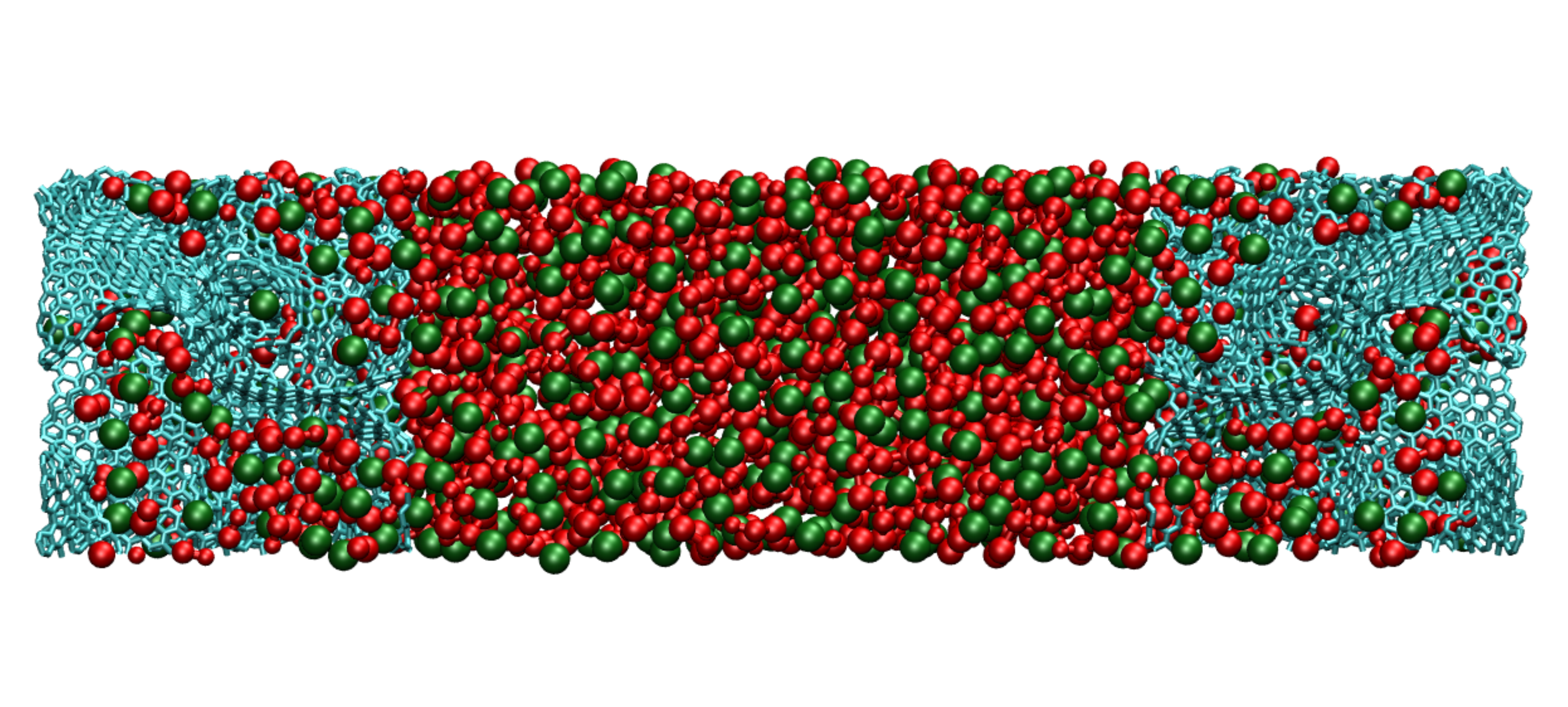}
\end{center}
\caption{Representative snapshot extracted from a simulation of a BMI-PF$_6$ ionic liquid and CDC nanoporous electrodes. Green spheres: PF$_6^-$, Red spheres: BMI$^+$, Blue rods: carbon-carbon bonds.}
\label{fig:snapshotporeux}
\end{figure*}

As soon as porous electrodes are considered, several additional effects arise. First of all, due to the confinement the ions \revision{close to the electrode surface} become less coordinated. In  recent work, we  performed simulations of a supercapacitor consisting of a BMI-PF$_6$ ionic liquid and carbide-derived carbons (CDCs) electrodes~\cite{merlet2012a}. CDCs are nanoporous carbons with well-defined pore size distribution~\cite{gogotsi2003a,dash2006a}, for which a strong increase of the (integral) capacitance \revision{with decreasing pore size was shown} experimentally for various electrolytes~\cite{chmiola2006a,largeot2008a,chmiola2008a,lin2009a}. In our simulations, we have used structures which were obtained by Palmer {\it et al.} using quenched molecular dynamics~\cite{palmer2010a} and were shown to
correspond to CDCs synthesized from crystalline titanium carbide using different chlorination temperatures. Figure \ref{fig:snapshotporeux} displays a snapshot extracted from one of our simulations of this system.  For a null voltage difference between the electrode\revision{s}, the coordination number drops from seven in the bulk to four inside the electrodes~\cite{merlet2012a}.


Then, as soon as a potential is applied between the electrodes, an exchange of
ionic species occurs between the electrolyte and \revision{those initially absorbed within} the nanoporous carbons. In an elegant
analytical study, Kondrat and Kornyshev have shown that this \revision{is facilitated by} the
image charges at the surface of the metallic electrode, which exponentially
screen the electrostatic interactions \revision{between  ions close to the surface}~\cite{kondrat2011a}. Our
numerical simulations are in qualitative agreement with this model\revision{.} The
volume occupied by the liquid inside the electrode remains almost constant, but
its composition is changed and it is no more electrically neutral \revision{; the net
charge at the surface of the electrode exactly balances that of the adsorbed
liquid}.   In addition, the coordination number of anions (resp.  cations)  inside the 
negative (resp. positive) electrode decreases even further, passing from 4 to 3, for an applied potential of 1~V. On the contrary, cations (resp. anions) become more highly coordinated inside the pore (their average coordination number is then around 5).    

\begin{figure}[ht]
\begin{center}
 \includegraphics[width=\columnwidth]{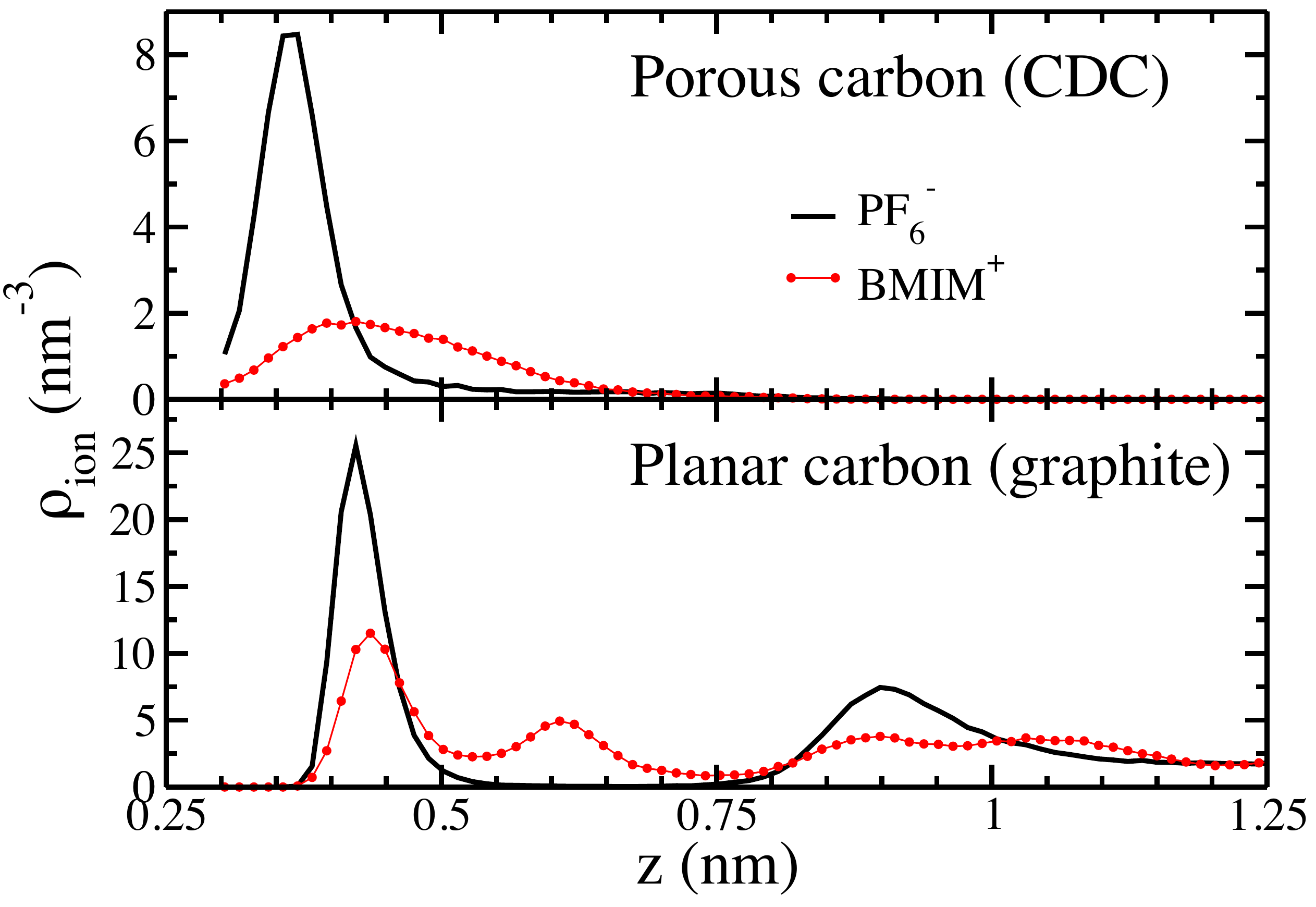}
\end{center}
\caption{Ionic density profiles for the BMI$^+$ and PF$_6^-$ ions adsorbed on graphite or porous carbon surfaces.}
\label{fig:dens-compporeuxgraphite}
\end{figure}

In agreement with experiments, the associated integral capacitance is then very
high: In our study, we calculated values of 87 and 125~F per gram of carbon for
two different CDC structures and applied potentials of 0.75--1~V, while it was of 30 F g$^{-1}$ in the case of \revision{planar}
graphite electrodes. In order to understand this difference, it is very useful
to compare the corresponding ionic density profiles. Figure
\ref{fig:dens-compporeuxgraphite} displays these profiles
for one of the porous carbon structure\revision{s} and for the planar graphite surface. Two
key differences are observed. Firstly, the ions approach closer to the surface
by 0.7~\AA\ \revision{in the porous electrode case}. This partly explains the increase of the capacitance, since a
shorter distance will give rise to a larger local polarization of the carbon
surface. Secondly, due again to the confinement inside the pore, only one layer
of ions is now adsorbed on the surface. As a consequence,
\revision{overscreening cannot take place} inside nanoporous carbon electrodes, an effect which is clearly visible when comparing the density of charge on the liquid side of the interface with the surface charge~\cite{merlet2012a}\revision{. Although in the case of graphite the charge in
the first adsorbed layer reaches a value
which is three times higher than the charge of the electrode itself, in the CDC the total charge in the first layer balances exactly that of the electrode; this} results 
in a much better efficiency for the charge storage.

Other recent simulation works have focused on the study of the adsorption of ionic liquids inside nanoporous electrodes. Idealized slit pores or regular carbon structures such as the interior of nanotubes were considered, either by molecular simulations or using a classical density functional theory approach~\cite{bishop2009a,yang2009a,shim2010a,vatamanu2010a,wu2011a,feng2011c,jiang2011a,kondrat2011b,pizio2012a,xing2013a}. All these works were in qualitative agreement with the increase of capacitance for pore sizes matching the ionic size observed experimentally and reproduced in figure \ref{fig:largeot}. Interestingly, several studies reported an oscillatory behavior of the capacitance with pore size~\cite{wu2011a,feng2011c,jiang2011a}, an effect which could not be observed experimentally because real electrodes used up to now always exhibit a pore size distribution. In addition, our simulations showed that due to the variety of environment\revision{s} encountered by the ions inside the relatively disordered porous carbons, the capacitance is not only linked to the size of the pores but also depends on the local structure inside the porosities~\cite{merlet2012a}, in agreement with the curvature and roughness effects described in the previous section. Finally, it is worth noting that the effect of pore size dispersity has been addressed in a single study so far by Kondrat et {\it al.}~\cite{kondrat2012a}.

\section{Concluding remarks}

While to date most theoretical work has been devoted to the simulation of pure ionic liquids, most experimental devices \revision{actually use} ions dissolved in a solvent. Aqueous electrolytes, which are used in batteries and display a high ionic conductivity, have been studied for a long time by experiments, theory and simulations. In fact, the understanding of charged solid/liquid interfaces has been the subject of over a century of theoretical work, from the early studies of Gouy and Chapman and the concept of double layer to the most recent molecular simulations. However aqueous electrolytes offer only a rather narrow electrochemical window and organic solvents such as acetonitrile (ACN) or propylene carbonate (PC) are usually preferred \revision{for EDLC applications}. Future theoretical work should thus focus on the solvation of organic ions in these liquids and at their interface with electrodes~\cite{merlet2013a,bhuiyan2013a}. The challenge will then be to describe how solvation impacts the structure of the electrolyte at the interface and the resulting charge storage process. 

Significant theoretical work should also focus on the solid counterpart of the electrolyte, namely the description of the porous electrode. While the importance of using realistic electrode structure to capture the microscopic charging process has been recently demonstrated~\cite{merlet2012a}, resort to simplified theories remains necessary to predict generic trends e.g. of the ratio between molecular and pore sizes or of polydispersity~\cite{kondrat2012a}. We thus anticipate that both routes will still provide useful insights. In the former case, the topological analysis of experimental microscopic structures should reveal characteristic properties which influence the capacitance beyond the average pore size or even the pore size distribution. This will improve the understanding of the intimate interplay between the porous structure and the electrolyte, in particular for the dynamical aspect of the charge and discharge. In the latter case, simplified studies will allow for a systematic screening for the design of improved materials.

From the theoretical point of view, it would also be desirable to improve the description of the electrode material. The shift from a constant charge to a constant potential \revision{within the classical simulation} framework has already been a significant step in the description of the electrode~\cite{merlet2013b} \revision{but it would be useful to have ways of systematically improving the description of the short-range interactions} between ions and the atoms of the electrode surface. We note that Pounds {\it et al}~\cite{pounds2009b} implemented a ``force-fitting" method to inject first-principles parameterization for a classical ion-electrode force-field of a prescribed functional form - but further systematic investigations are necessary to expose what physical effects  should be included in such models, and {\it ab initio} simulations of small-scale systems would be useful. We also note that the currently available constant potential methods treat the electrode as an ideal metal. It is not {\it a priori} obvious that this is appropriate for a disordered metal ~\cite{pastewka2011a} or porous carbon (or whether there are limitations on the appropriate length- or time-scales on which it might be justified), and it is certainly not appropriate for the semiconductor surfaces which one would like to be able to treat.

\revision{ Several further directions} are likely to be explored in the near future within the framework of the currently available methods. The most immediate step, even though not straightforward, is to improve the numerical efficiency of the algorithms to maintain a constant potential condition. Such phenomena as electrowetting at metallic surfaces~\cite{millefiorini2006a} or the distribution of ions around particular topological features, like asperities or step-edges are ripe for exploration with the constant potential method. The current review has been confined to non-faradaic process at the electrodes surface, but charge transfer induced by a change in the electrode potential between the electrode and redox species in the electrolyte has been investigated~\cite{reed2008a} with the constant potential method within the framework of Marcus Theory. This work has shown how to relate, at the atomistic level, the macroscopic potential applied to the cell to potentials felt by the redox species in the electrolyte.

\section*{Acknowledgements}
 We are grateful to Alexei Kornyshev, Patrice Simon, Pierre-Louis Taberna, Ren\'e van Roij and Susan Perkin for useful discussions. We acknowledge the support of the French Agence Nationale de la Recherche (ANR) under grant ANR-2010-BLAN-0933-02 (`Modeling the Ion Adsorption in Carbon Micropores'). We are grateful for the computing resources on JADE (CINES, French National HPC) obtained through the project x2012096728. This work made use of the facilities of HECToR, the UK's national high-performance computing service, which is provided by UoE HPCx Ltd at the University of Edinburgh, Cray Inc and NAG Ltd, and funded by the Office of Science and Technology through EPSRC's High End Computing Programme.

\end{document}